\documentclass[11pt,dvips]{article}

\usepackage{epsfig,times}
\usepackage{picinpar}
\usepackage{wrapfig}
\usepackage{floatflt}
\makeatletter
\renewcommand{\section}{\@startsection{section}{1}{0in}
	{0.4\baselineskip}{0.1\baselineskip}{\Large\bf}}
\renewcommand{\subsection}{\@startsection{subsection}{2}{0in}
	{0.25\baselineskip}{-\baselineskip}{\large\bf}}
\renewcommand{\subsubsection}{\@startsection{subsubsection}{3}{0in}
	{0.1\baselineskip}{-\baselineskip}{\normalsize\bf}}
\makeatother
\begin{document}

%
\thispagestyle{myheadings}
%
\markright{OG 2.2.21}
\begin{center}
%
{\LARGE \bf TeV Measurements of Young Pulsars and Supernova Remnants}
\end{center}


\begin{center} 

{\bf P.M.~Chadwick, K.~Lyons, T.J.L.~McComb, K.J.~Orford,
M.G.G.~O'Connell, J.L.~Osborne, S.M.~Rayner, S.E.~Shaw, and
K.E.~Turver}\\ {\it Department of Physics, Rochester Building, Science
Laboratories, University of Durham, Durham, DH1~3LE, U.K.} \end{center}


\begin{center} {\large \bf Abstract\\} \end{center}
\vspace{-0.5ex} 

Observations have been made with the University of Durham Mark 6
telescope of a number of supernova remnants and young pulsars (Vela
pulsar, PSR B1055--52, PSR J1105--6107, PSR J0537--6910 and PSR
B0540--69). No VHE gamma ray emission, either steady or pulsed, has been
detected from these objects. %

\vspace{1ex}

%
%

\section{Introduction}

The {\em Compton Gamma Ray Observatory} ({\em CGRO}) telescopes have
detected pulsed gamma radiation from 7 pulsars so far: Crab, Vela,
Geminga, PSR B1509--58, PSR B1706--44, PSR B1951+32 and PSR B1055--52
(see e.g. Thompson et al. 1997). Of these, the Crab and PSR B1706--44
are confirmed VHE gamma ray emitters, and the Vela remnant has been
detected by the CANGAROO group. Although the gamma ray emission from all
the pulsars detected with the EGRET telescope is pulsed, thus far no
imaging VHE gamma ray telescope has detected pulsed radiation at TeV
energies from any of the EGRET pulsars. 

We have previously presented limits on VHE gamma ray emission from a
number of Southern hemisphere pulsars using the University of Durham
Mark 3 telescope (Brazier et al. 1990). We present here the results of
VHE gamma ray observations of five plerions using the Mark 6 imaging
telescope; two EGRET sources (Vela and PSR B1055--52) and three X-ray
emitting pulsars, PSR J1105--6107, PSR J0537--6910 and PSR B0540--69. We
have searched for both steady and pulsed emission from these objects.

\section{Observations}

The Durham University Mark 6 telescope is described in detail elsewhere
(Armstrong et al., 1999). It consists of three 7 m diameter parabolic
flux collectors mounted on a single alt-azimuth platform. A 109-element
imaging camera is mounted at the focus of the central mirror, with
low-resolution cameras each consisting of 19 pixels mounted at the focus
of the outer (left and right) flux collectors. These detectors operate
as a 4-fold temporal plus 3-fold spatial triggering system, which
provides for a robust muon-free trigger and a low threshold energy
($\geq 300 - 400$ GeV).

Data from all objects except the PSR J0537--6910/PSR B0540--69 field
were taken in 15-minute segments. Off-source observations were taken by
alternately observing regions of sky which differ by $\pm 15$ minutes in
RA from the position of the object to ensure that on- and off-source
segments have identical zenith and azimuth profiles. Data were accepted
for analysis only if the sky was clear and stable and the gross counting
rates in each on-off pair were consistent at the $2.5 \sigma$ level. In
the case of PSR J0537--6910 and PSR0540--69, the two objects were
tracked and kept in the field of view at all times during the
observations. 

\subsection{Vela pulsar}

The Vela pulsar is comparatively close to Earth ($\sim 500$ pc distant)
and so the surrounding nebula is well studied. X-ray studies show that
there is an X-ray jet and so there may be evidence of a pulsar wind
(Markwardt \& \"{O}gelman 1995). Observations with {\em ASCA} have
suggested that the jet emission is non-thermal (Markwardt \& \"{O}gelman
1997), which leads to the strong possibility of the production of
Compton-boosted VHE gamma rays. Extensive observations of the Vela
pulsar have been made with EGRET (Fierro et al. 1998). The light curve
is double peaked with emission occuring in the phase interval between
the two peaks. There is evidence for some weak unpulsed emission ($4.4\%
\pm 0.9\%$ of the total emission) and a spectral turnover is seen at
about 1 GeV.

A number of limits to pulsed VHE emission from the Vela pulsar have been
obtained using first generation non-imaging detectors. The CANGAROO
group have detected DC VHE emission using the 3.8m imaging telescope
obtaining a flux of $(2.9 \pm 0.5 \pm 0.4) \times 10^{-12} {\rm~cm}^{-2}
{\rm~s}^{-1}$ above $2.5 \pm 1.0$ TeV (Yoshikoshi et al. 1997) from a
region offset from the Vela pulsar position by about $0.13^\circ$. They
find that the emission was steady over a baseline of 2 years and that
there was no evidence for pulsed emission.

\begin{table}[t!]
\begin{center}
\begin{tabular}{@{}lcclcc} \hline \hline
Object & Date & No. of &Object & Date & No. of\\
& & ON source scans & & & ON source scans \\ \hline
Vela pulsar & 1996 Apr 14 & 1 & PSR J1105--6107 & 1997 Mar 31 & 5 \\
Vela pulsar & 1996 Apr 18 & 4 & PSR J1105--6107 & 1997 Apr 1 & 6 \\
Vela pulsar & 1996 Apr 19 & 4 & PSR J1105--6107 & 1997 Apr 3 & 9 \\
Vela pulsar & 1996 Apr 20 & 5 & PSR J1105--6107 & 1997 Apr 4 & 9 \\
Vela pulsar & 1996 Apr 21 & 5 & PSR J1105--6107 & 1997 Apr 5 & 7 \\
Vela pulsar & 1996 Apr 22 & 3 & PSR J1105--6107 & 1997 Apr 6 & 9 \\
Vela pulsar & 1997 Feb 6 & 7 & PSR J1105--6107 & 1997 Apr 7 & 4 \\
Vela pulsar & 1996 Feb 7 & 6 & PSR J1105--6107 & 1997 Apr 8 & 2 \\
LMC pulsars & 1998 Mar 21 & 110 mins & PSR J1105--6107 & 1997 Apr 9 & 9 \\
LMC pulsars & 1998 Mar 22 & 130 mins & PSR J1105--6107 & 1997 Apr 10 & 7 \\
LMC pulsars & 1998 Mar 24 & 100 mins & PSR B1055--52 & 1996 Mar 19 & 5 \\
LMC pulsars & 1998 Mar 27 & 90 mins & PSR B1055--52 & 1996 Mar 20 & 5 \\
LMC pulsars & 1998 Mar 28 & 80 mins & & & \\ \hline
\end{tabular}
\end{center}
\caption{Observing log for observations of pulsars made with the
University of Durham Mark 6 telescope. The number of 15 minute ON-source
scans obtained is shown, except for the LMC pulsars where the total
exposure time is shown (see text).}
\label{observing_log} 
\end{table}

\subsection{PSR B1055--52}

The radio pulsar PSR B1055--52 was discovered in 1972 by Vaughan \&
Large (1972). It has a pulse period of 197.11 ms and a characteristic
age of 530 kyr. It has been shown (McCulloch et al. 1976) to be in the
small class of radio pulsars with a strong interpulse half a cycle away
from the main pulse, and as such has a radio light curve similar to that
of the Crab pulsar. It was detected as a soft X-ray source by the {\em
Einstein Observatory} (Cheng \& Helfland 1983) and later by {\em EXOSAT}
(Brinkman \& \"{O}gelman 1987). Neither the {\em Einstein} nor the {\em
EXOSAT} observation provided evidence for pulsed X-ray emission, but
later observations made with the the {\em ROSAT} telescope showed
evidence for pulsed X-ray emission (\"{O}gelman \& Finley 1993). The
EGRET detector on board {\em CGRO} discovered high energy gamma ray
emission from PSR B1055--52 (Fierro et al. 1993). More extensive
observations (Thompson et al. 1998) have shown that the gamma ray
emission is complicated. The source has a complex light curve with no
detectable unpulsed emission. The gamma-ray energy spectrum is flat,
with no evidence for a cut off at energies up to 4 GeV.

\subsection{PSR J1105--6107}

PSR J1105--6107 was discovered with the Parkes radio telescope in 1994,
with follow-up timing observations being made over the next two years
(Kaspi et al. 1997). Its pulse period is 63 ms and it is young, having a
characteristic age of 63 kyr. The pulsar is located close to the
supernova remnant MSH 11--61A, which may be associated with the pulsar.
It is also positionally coincident with the EGRET source 2EG
J1103--6106, and if the pulsar is associated with the EGRET object then
the observed $\gamma$-ray flux suggests an efficiency for conversion of
spin-down luminosity to $\gamma$-rays of approximately 3\%. Significant
X-ray emission was detected from PSR J1105--6107 using the {\em ASCA}
observatory (Gotthelf and Kaspi 1998) and {\em RXTE} (Steinberger, Kaspi
\& Gotthelf 1998). The X-ray emission shows no evidence for pulsations,
and it is suggested that the X-ray emission originates in a
pulsar-powered synchrotron nebula. 

\subsection{PSR J0537--6910 and PSR B0540--69}

These two fast pulsars are within the Large Magellanic Cloud and can be
observed simultaneously with the Mark 6 telescope.

PSR J0537--6910 is a fast (16 ms) pulsar embedded in the supernova
remnant N157B in the Large Magellanic Cloud. The pulsar was detected in
{\em RXTE} data by Marshall et al. (1998). The supernova remnant has
been suggested to be a Crab-like plerion from its central peaked
morphology and flat spectra in both radio and X-ray (Wang \& Gotthelf
1998). No pulsed radio emission has been detected (Crawford et al.
1998).

PSR B0540--69 is a 50 ms pulsar in the Large Magellanic Cloud. It was
first detected in X-ray data obtained with {\em Einstein} (Seward,
Harnden \& Helfand 1984) and is embedded in a young SNR. Optical
pulsations have been detected (Middleditch and Pennypacker 1985) but
only weak radio emission has been observed (Manchester et al. 1993).
Again, the SNR is a synchrotron nebula and the system is Crab-like.

\section {Data Analysis}

Data reduction and analysis followed our standard procedure, which has
been described in detail previously (Chadwick et al. 1999). The
selection criteria applied to the data are summarized in Table 2; these
are a standard set of criteria developed from our successful
observations of PKS 2155--304, and allow for the variation of image
parameters with image size.

To check for the presence of a pulsed signal, the phase of each event
was evaluated using the ephemeris nearest the observation date from the
Princeton database or other published sources. For data from the Vela
pulsar, PSR B1055--52 and PSR J1105--6107 the events were then binned in
20 phase bins. No single bin showed a significant excess compared with
the mean. Rayleigh and $\chi^2$ tests were performed on the binned data.
No light curve showed significant Rayleigh power. The pulsed flux limit
quoted is the limit to 5\% duty cycle emission implied by the lack of a
significant excess in any bin. The data from PSR J0537--6910 and PSR
B0540--69, for which no suitably accurate epehemerides were available,
were subjected to a Rayleigh test over a small range of periods about
the most likely period. No significant Rayleigh power was detected at
any period.

\section{Results}

\begin{table}[b]
\begin{center}
\begin{tabular}{@{}lcccc} \hline \hline
Object & Estimated & Flux Limit (DC) & Flux Limit (pulsed) & Ephemeris \\
& Threshold (GeV) & $ (\times 10^{-10} {\rm~cm}^{-2} {\rm~s}^{-1})$ & $ (\times 10^{-11} {\rm~cm}^{-2} {\rm~s}^{-1})$ & reference \\ \hline
Vela pulsar & 300 & 0.50 & 1.3 & [1] \\
PSR B1055--52 & 300 & 1.3 & 6.8 & [2] \\
PSR J1105--6107 & 400 & 0.22 & 0.53 & [3] \\
PSR J0537--6910 & 400 & 0.61 & 1.0 & [4] \\
PSR B0540--69 & 400 & 0.61 & 1.1 & [5] \\ \hline
\end{tabular}
\end{center}

\caption{Flux limits for observations of pulsars made with the
University of Durham Mark 6 Telescope. Refs. [1] Arzoumanian et al. 1992
GRO/radio timing data base, unpublished; [2] Kaspi et al. 1996,
unpublished; [3] Kaspi et al. 1997, Ap. J., 485, 820; [4] Wang \&
Gotthelf 1999 ApJ, 509, L109; [5] Deeter et al. 1999, ApJ, 512, 300.}

\label{results_table} 
\end{table}

The dataset for each source has been tested for the presence of both
pulsed and steady gamma ray signals as described above. In addition, as
the Vela SNR source reported by the CANGAROO group is $\sim 0.13^\circ$
from the pulsar position, a false source analysis has been performed for
this object. The threshold energy for the observations has been
estimated on the basis of preliminary simulations, and is in the range
300 to 400 GeV for these objects, depending on the elevation at which
observations were made. The collecting areas which have been assumed are
$5.5 \times 10^{8}~{\rm cm}^{2}$ at an energy threshold of 300 GeV and
$1 \times 10^{9}~{\rm cm}^{2}$ at 400 GeV. These are subject to
systematic errors estimated to be $\sim 50\%$. We have assumed that our
current selection procedures retain $\sim 20\%$ of the $\gamma$-ray
signal. We have no evidence for the emission of either steady or pulsed
VHE gamma rays from any of the plerions, and the flux limits are
summarized in Table \ref{results_table}. All steady flux limits are
$3~\sigma$ limits, based on the maximum likelihood ratio test (Gibson et
al. 1982). Pulsed flux limits for PSR J0537--6910 and PSR B0540--69 are
based on the percentage pulsed flux which would be required to produce a
$3~\sigma$ pulsed detection using the Rayleigh test. The pulsed flux
limits for the other objects are based on the pulsed flux that would be
required to yield a $3~\sigma$ excess in any single bin of a 20 bin
lightcurve.

\section{Discussion}

The only object considered in this paper which has been detected at TeV
energies is the Vela pulsar/nebula. An extrapolation of the flux
detected with the CANGAROO telescope at $2.5 \pm 1.5$ TeV to our
threshold energy of about 300 GeV suggests that we might expect to have
detected the offset source described by Yoshikoshi et al. (1997).
However, taking into account the errors on our flux and threshold energy
estimates, CANGAROO's flux and energy threshold estimates and the errors
on the measured spectral index, we find that our result is not
incompatible with that of Yoshikoshi et al. (1997).

PSR B1055--52 is a topical object, with the recent publication of the
detailed analysis of the EGRET results (Thompson et al., 1998).
Extrapolating the EGRET spectrum to TeV energies indicates an expected
flux which is very close to our flux limit. If further observations do
not yield a detection of this object at 400 GeV, then we may have
evidence for a break in the power law spectrum.

Both PSR J0537--6910 and PSR B0540--69 are good candidates for TeV
emission on the basis of their radio and X-ray characteristics. However,
their distance from the earth (they are both situated in the LMC) means
that considerable further exposure is likely to be necessary to detect a
significant flux.

We are grateful to the UK Particle Physics and Astronomy Research
Council for support of the project. 
\vspace{1ex}
\begin{center}
{\Large\bf References}
\end{center}
%
Armstrong, P., et al. 1999, Exp. Astro., in press\\
Brazier, K. T. S., et al. 1990, 21st ICRC (Adelaide), 2, 304\\
Brinkmann, W. \& \"{O}gelman, H. B. 1987, A\&A, 182, 71\\
Chadwick, P. M., et al. 1999, ApJ, 513, 161\\
Cheng, A. F. \& Helfand, D. J. 1983, ApJ, 271, 271\\
Crawford, F., et al. 1998, astro-ph/9808358\\
Fierro, J. M., et al. 1995, ApJ, 413, L27\\
Fierro, J. M., Michelson, P. F., Nolan, P. L., \& Thompson, D. J. 1998, ApJ, 494, 734\\
Gibson, A. I., et al. 1982, Proc. Intl. Workshop on Very High Energy
Gamma Ray Astro., Bombay: Tata Institute, ed. P. V. Ramana Murthy \& T.
C. Weekes, p. 97\\
Gotthelf, E. V. \& Kaspi, V. M. 1998, ApJ, 497, L29\\
Kaspi, V. M., et al. 1997, ApJ, 485, 820\\
Manchester, R. N., et al. 1993, ApJ, 403, L29\\
Markwardt, C. B. \& \"{O}gleman, H. B. 1995, Nature, 375, 40\\
Markwardt, C. B. \& \"{O}gleman, H. B. 1997, ApJ, 480, L13\\
Marshall, F. E., Gotthelf, E. V, Zhang, W., Middleditch, J. \& Wang, Q. D. 1998, ApJ, 499, L179\\
McCulloch, P. M., Hamilton, P. A., Ables, J. G. \& Komesaroff, M. M. 1976, MNRAS, 175, 71P\\
\"{O}gelman, H. B. \& Finley, J. P. 1993, ApJ, 413, L31\\
Steinberger, J., Kaspi, V. M. \& Gotthelf, E. V., astro-ph/9809367\\
Thompson, D. J., Harding, A. K., Hermsen, W. \& Ulmer, M. P. 1997, Proc.
Fourth Compton Symposium, ed. C. D. Dermer, M. S. Strickman \& J. D.
Kurfess, AIP Conf Proc. 410, p. 39\\
Thompson, D. J., et al. 1998 astro-ph/9811219\\
Vaughan, A. E. \& Large, M. I. 1972, MNRAS, 156, 27P\\
Wang, Q. D. \& Gotthelf, E. V. 1998, ApJ, 494, 623\\
Yoshikoshi, T., et al. 1997, ApJ, 487, L65\\
\end{document}